\documentclass[pre,preprint]{revtex4}
\usepackage{graphicx}
\usepackage{amsmath}

\begin{document}

\title{The looping probability of random heteropolymers helps to understand the scaling properties of biopolymers}
\author{Y. Zhan}
\affiliation{Friedrich Miescher Institute for Biomedical Research, Maulbeerstrasse 66, CH-4058 Basel, Switzerland}
\author{L. Giorgetti}\email{luca.giorgetti@fmi.ch}
\affiliation{Friedrich Miescher Institute for Biomedical Research, Maulbeerstrasse 66, CH-4058 Basel, Switzerland}
\author{G. Tiana}\email{guido.tiana@unimi.it}
\affiliation{Center for Complexity and Biosystems and Department of Physics, Universit\`a degli Studi di Milano and INFN, via Celoria 16, 20133 Milano, Italy}
\date{\today}

\begin{abstract}
Random heteropolymers are a minimal description of biopolymers and can provide a theoretical framework to the investigate the formation of loops in biophysical experiments.
A two--state model provides a consistent and robust way to study the scaling properties of loop formation in polymers of the size of typical biological systems. Combining it with self--adjusting simulated--tempering simulations, we can calculate numerically the looping properties of several realizations of the random interactions within the chain. Differently from homopolymers, random heteropolymers display at different temperatures a continuous set of scaling exponents. The necessity of using self--averaging quantities makes finite--size effects dominant at low temperatures even for long polymers, shadowing the length--independent character of looping probability expected in analogy with homopolymeric globules. This could provide a simple explanation for the small scaling exponents found in experiments, for example in chromosome folding.
\end{abstract}

\maketitle

%%%%%%%%%%%
\section{Introduction}
%%%%%%%%%%%

Most of biological molecules are polymers, and the formation of contacts between monomers which are not close along the chain usually plays an important biological role. For example, in the chromatin fibre the approaching of an enhancer to a promoter located millions of bases away is often necessary to trigger transcription \cite{Blackwood:1998}. In the case of proteins, the formation of non--covalent interactions between distant amino acids is, in many cases, among the first steps in the folding process \cite{Bruun:2010}. 

There are several experimental techniques to study, either directly or indirectly, the formation of contacts between pairs of monomers as a function of their distance $N$ along the polymeric chain. Arguably, when $N$ is large enough, the detailed chemistry of the system looses importance and one can highlight its more general physical properties. In the case of chromosome folding, it was found by Hi--C experiments that  the binding probability between chromosomal loci depends on $N$  as a power law $N^{-\beta}$ with exponent $\beta \approx 1$ above the megabase-scale in human chromosomes\cite{Lieberman:2009} and even lower at a smaller scale \cite{Sanborn:2015}. The looping probability of peptides with repeated AGQ sequence, measured by FRET,  displays a power--law with exponent 1.55 in water and 1.7 in urea and guanidine \cite{Buscaglia:2006}. The folding rate of proteins, measured by stopped-flow experiments, was shown to correlate with the (rescaled) average value of $N$ of pairs of amino acids which are in contact in the native state \cite{Plaxco:1998}. In long RNA chains the contact probability displays an exponent $\beta\approx 1$ \cite{Liu:2006}.

The simplest theoretical framework to describe the contact formation in a biopolymer at equilibrium as a function of $N$ is that of two interacting monomers linked by a homopolymer. One can employ a two--state description of the system, assuming that the formation of the contact between the two ends does not change the density of the polymer. In this case, if $\epsilon<0$ is the energy gain of the system upon formation of the contact, the associated probability can be approximated  as
\begin{equation}
c(N)=\frac{\exp(-\epsilon/T)}{g(N)+\exp(-\epsilon/T)},
\label{eq:twostate}
\end{equation}
where $g(N)$ is the density of state of the system displaying the contact with respect of the unbound state. Its shape depends on the properties of the linking homopolymer. If this can be regarded as an ideal chain then $g(N)=N^{3/2}$, if it is a random coil due to the repulsion between its elements, $g(N)=N^{9/5}$, while it is  constant in a globule \cite{deGennes:book}. In the limit of large $N$ one then expects a scaling law of the  type $c\propto N^{-\beta}$, with $\beta=0$, $1/2$ or $9/5$, as discussed above. 
The scaling exponents found for repeat peptides \cite{Buscaglia:2006} lie between those expected for an ideal chain and a random coil.
In the case of chromatin, the anomalous exponent $\beta=1$ found in experiments is not compatible with the above model and was explained either with non-equilibrium effects \cite{Mirny:2011,Grosberg:1988},  with interactions mediated by floating molecules \cite{Barbieri:2012} or by energy--driven mechanisms \cite{Goloborodko:2015,Sanborn:2015}.

However, in most cases, the monomers which build polymers of biological interest are chemically heterogeneous, and the homopolymeric assumption is questionable. The problem we would like to address in the present work is the role of heterogeneous interactions in determining the scaling properties of the contact probability between monomers. Specifically, we study the looping probability of random heteropolymers \cite{Shakhnovich:1989}, regarding them as a minimal model for biomolecules.

To investigate this problem, we made use of a simple model, in which the polymer is described as a chain of beads connected by rigid links. Pairs of beads interact through a spherical--well potential with a hard--core of radius $r_H$, a width $r$ and a depth $B_{ij}$ which depends on the specific pair. For sake of generality, we considered the energies $B_{ij}$ as quenched stochastic variables, defined by a Gaussian distribution of mean $B_0$ and standard deviation $\sigma_B$. In this way we are not focusing on a particular kind of biopolymer, but we are looking for the general properties which arise only because of the heterogeneity of the interactions. 

Operatively, we investigated the equilibrium contact probability of heteropolymeric chains by mean of numerical simulations. In this case, the stochasticity of the interaction energies is modelled generating several realizations of the set of Gaussian variables, and for each of them carrying out a conformational sampling. This approach poses the problem of averaging the results of the samplings over the quenched energies. The contact probability itself does not result to be a self--averaging quantity, and consequently its average over the realizations of the quenched variables $B_{ij}$ is poorly informative  \cite{Lifshits:1942}. In Sect. \ref{sect:selfav} we discuss under which conditions the average of quantities associated with the contact probability are informative.

\begin{figure}
\includegraphics[width=\linewidth]{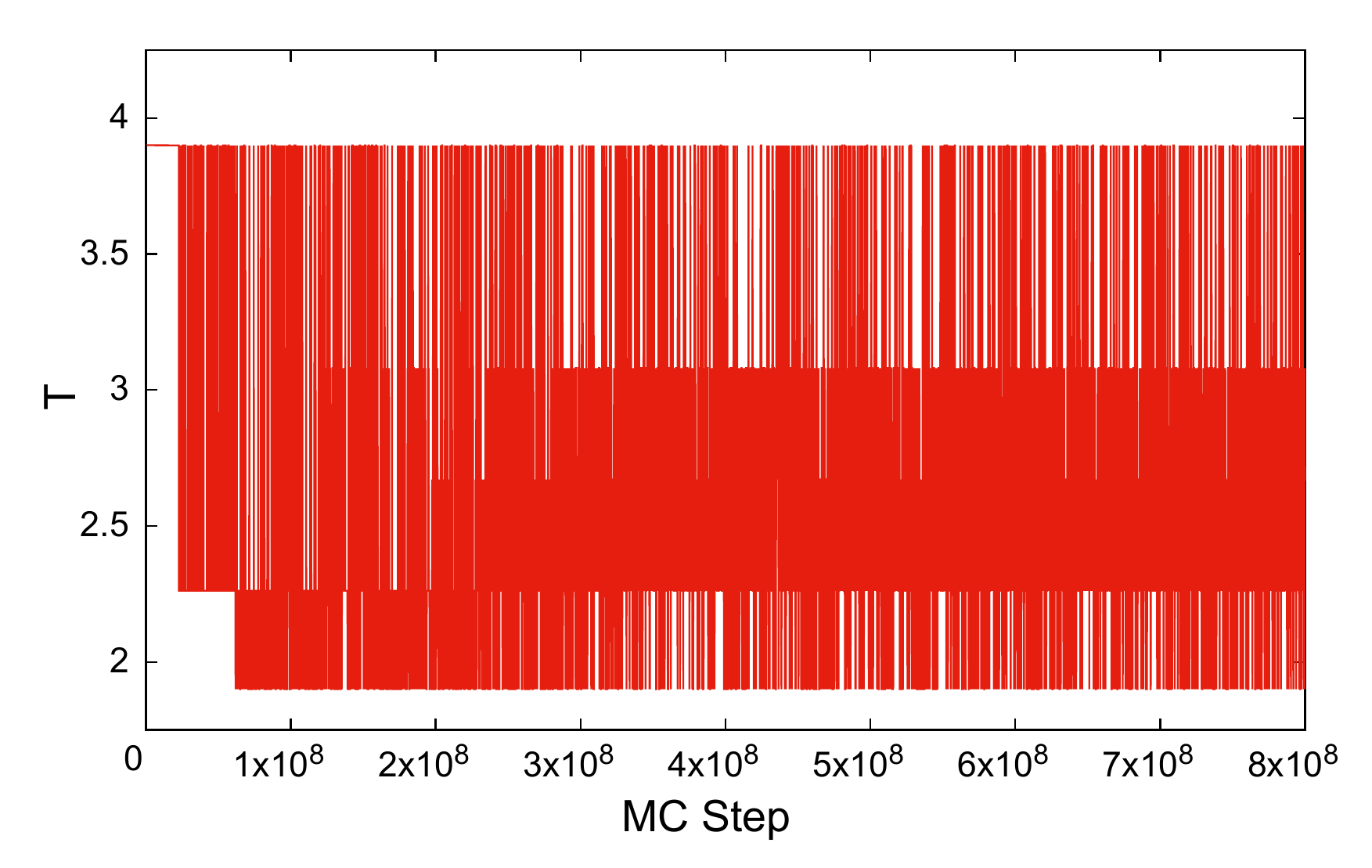}
\caption{(Color online) An example of evolution of the temperatures in the self--adjusting simulated tempering simulation.}
\label{fig:temperatures}
\end{figure}

Another problem one has to face is that the conformational sampling of disordered systems is computationally cumbersome, due to the roughness of the associated energy landscape. There are several computational techniques based on the multi--canonical ensemble which, sampling the system simultaneously at different temperatures, facilitate conformational sampling \cite{Swendsen:1986,Marinari:1992}. However, they rely on the choice of a set of temperatures optimized to enhance diffusion in temperature space. This set is not self--averaging, and consequently requires a manual fine tuning for each realization of the quenched variables. This is impractical if one wants to collect results from enough replicas to calculate reliable averages. To solve this problem in an automatic way, we made use of an adaptive simulated--tempering scheme developed in ref. \cite{Tiana:2011}. Starting from a high temperature, this algorithm performs a set of simulated--tempering samplings, adding iteratively lower temperatures, which are optimized self--consistently. An example of this procedure results in a sampling of different temperatures as that displayed in Fig. \ref{fig:temperatures}, which allows to calculate equilibrium averages of polymers up to $\sim 10^2$ monomers.

From the study of the looping probabilities of heteropolymers of different length, correctly averaged, we obtained the scaling behaviour as a function of the average interaction $B_0$ and the temperature $T$, setting $\sigma_B=1$ as reference energy. In the calculations, we chose \cite{Giorgetti:2014} $r_H=0.6$ (in units of the length of the link between consecutive beads), $r=1.5$ and a contact between the ends of the chain is defined if they are closer than $r$.

%%%%%%%%%%%%%%%%%
\section{The theoretical framework} 
\label{sect:theo}
%%%%%%%%%%%%%%%%%

In order to find the most appropriate way of calculating the scaling properties of the looping probability of a random heteropolymer, one can use a two--state model. One can assume that the bound and unbound states display, respectively,  $E_1+\epsilon$ and $E_2$, where 
$E_1$ and $E_2$ are quenched random variables regarded as the sum of the internal contact energies of the chain, while $\epsilon$ is the interaction energy between the ends of the chain. Further assuming that $E_1$ and $E_2$ are uncorrelated and that the two states have the same density, the central--limit theorem suggests that
\begin{equation}
p(E_1)=p(E_2)=\frac{1}{\sqrt{2\pi N\sigma^2}}\exp\left[-\frac{(E_{1,2}-N\epsilon_0)^2}{2N\sigma^2}\right],
\label{eq:gauss}
\end{equation}
where $N$ is the length of the chain, $\epsilon_0$ the average interaction between the monomers and $\sigma$ their standard deviation.
We define $\Delta E\equiv E_1-E_2$ and assume a density of states of the unbound state with respect to the looped state in the form of a power law of the kind $N^\beta$.  Thus, the entropy difference is $\beta\log N$ and the free energy difference between the two states is given by 
\begin{equation}
\Delta F=\Delta E+\epsilon+T\beta\log N
\label{eq:cN}
\end{equation}
where $\Delta E$ is a stochastic variable with distribution
\begin{equation}
p(\Delta E)=\frac{1}{\sqrt{4\pi N\sigma^2}}\exp\left(-\frac{\Delta E^2}{4N\sigma^2}\right).
\label{eq:pE}
\end{equation}
According to this model, the variability of the looping free energy, and consequently of the looping probability, at a given value of $N$ is due to the variability of the internal energy difference $\Delta E$. In other words, $\Delta E$ plays the role of the quenched disorder affecting the looping free energy defined as a function of $N$. The associated probability can be obtained inverting Eq. (\ref{eq:cN}) and substituting it into Eq. (\ref{eq:pE}), that is
\begin{equation}
p(\Delta F)=\frac{1}{\sqrt{4\pi N\sigma^2}}\exp\left(-\frac{\left(\Delta F - T\beta\log N -\epsilon \right)^2}{4N\sigma^2}\right).
\label{eq:pf}
\end{equation}
This probability can be maximized with respect to $\beta$ and $\epsilon$ according to a maximum-likelihood principle, obtaining
\begin{equation}
\beta=-\frac{1}{T}\frac{\sum_N \frac{1}{N}\cdot\sum_N \frac{\log(N)\Delta F}{N}-\sum_N \frac{\log(N)}{N}\cdot\sum_N \frac{\Delta F}{N}}{\sum_N \frac{1}{N}\cdot\sum_N \frac{\log^2(N)}{N}-\left(\sum_N \frac{\log(N)}{N}\right)^2},
\label{eq:beta}
\end{equation}
formally identical to the expression of a weighted linear regression.

From the simulations (or from a set of experiments) one can calculate the free energy difference $\Delta F$ from the contact probability
 \begin{equation}
\Delta F=-T\log\left[ \frac{c}{1-c}\right],
\label{eq:df}
\end{equation}
and use Eq. (\ref{eq:beta}) to obtain $\beta$ from a linear regression of $F$ versus $\log N$ with weights $N^{-1}$. This weighting is a consequence of the extensivity of the energy of the chain and has as  consequence that larger--$N$ points contribute less to the determination of $\beta$.

%%%%%%%%%%%%%%%%%
\section{The self--averaging issue} 
\label{sect:selfav}
%%%%%%%%%%%%%%%%%

The average $\overline{x}$ of a conformational property $x$ of the random heteropolymer over the quenched stochastic energies provides valuable information only if the associated standard error $\sigma_x$ is small, namely if the quantity is self--averaging \cite{Lifshits:1942}. In the thermodynamic limit, this corresponds to the condition 
\begin{equation}
\xi_x\equiv\frac{\sigma_x}{|\overline{x}|}\to 0.
\end{equation}

\begin{figure}
\includegraphics[width=\linewidth]{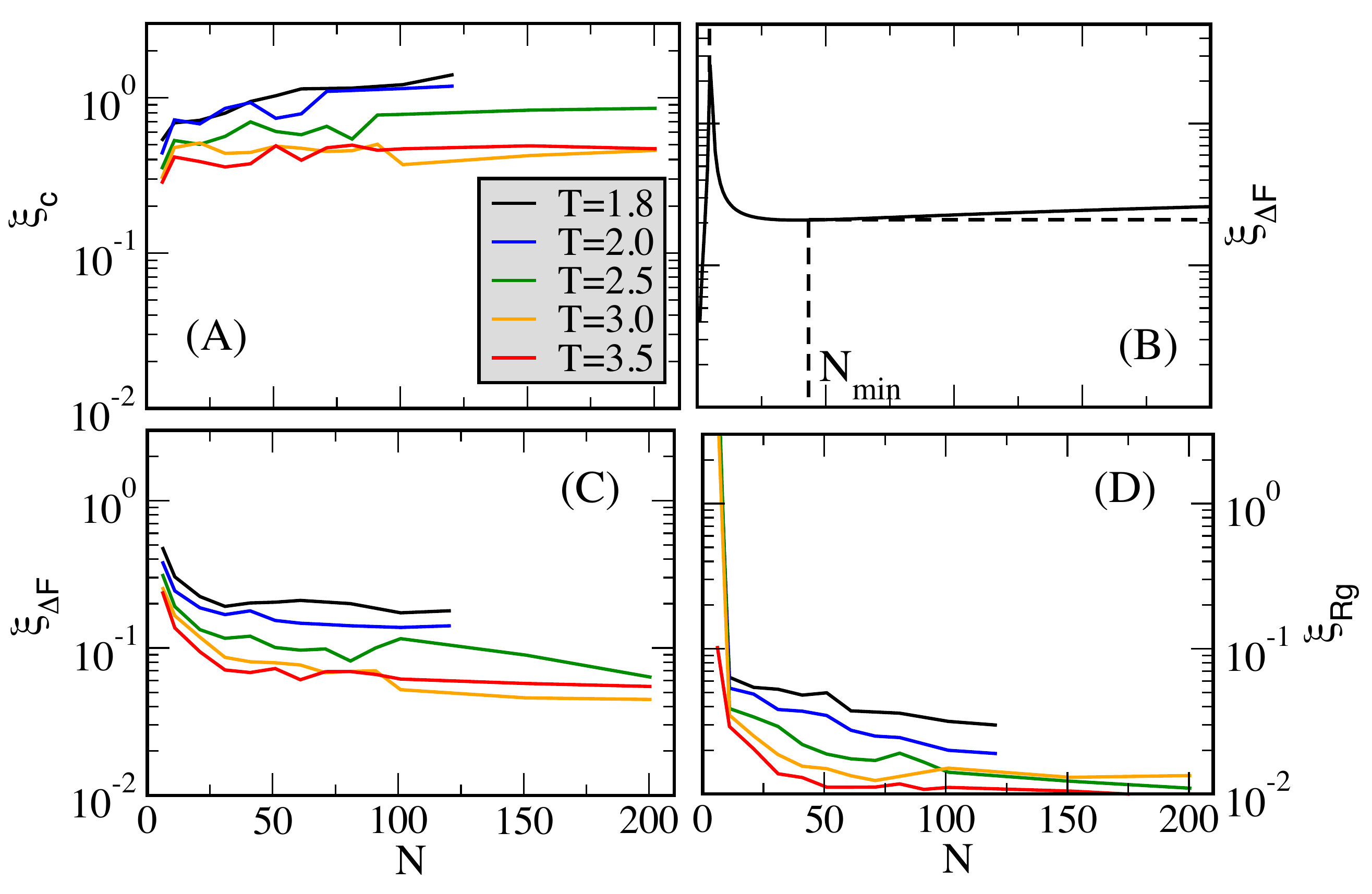}
\caption{(Color online) (A) The relative error $\xi_c$ associated with $c$; (B) a sketch of the theoretical behaviour of $\xi_{\Delta F}$ according to Eq. (\protect\ref{eq:xitheor}); (C) and (D) the relative error $\xi$ calculated for  $\Delta F$ and for the gyration radius $R_g$, respectively.}
\label{fig:selfav}
\end{figure}

Usually extensive properties are self--averaging \cite{Brout:1959}, while intensive properties, probability distributions and partition functions are not. Thus, we do not expect $c(N)$ to be self--averaging, and in fact $\xi_c$ is quite large, increasing above 1 quite fast as a function of $N$ at low temperatures (cf. Fig. \ref{fig:selfav}A). This is the reason why in the context of disordered systems one focuses the attention on free energies. However, in the present case we are considering a free--energy difference between two states of the system, which is expected to scale as $\log N$ according to Eq. (\ref{eq:cN}). The associated self--averaging parameter thus scales as $\chi_{\Delta F}\sim N^{1/2}/\log N$, which has a non--monotonic behaviour as a function of $N$, eventually diverging in the thermodynamic limit, although not very fast (cf. Fig. \ref{fig:selfav}B).

Thus, strictly speaking, $\Delta F$ is not self--averaging. Nor it is any quantity which can be derived by the contact probability $c$. However, if one is interested in finite systems of the typical size of biopolymers, a sufficient request is that the variability of $\Delta F$ associated with the disorder is smaller than its average, that is $\xi_{\Delta F}\ll 1$ in a specified interval of $N$.

Equation (\ref{eq:pf}) suggests that the variability of $\Delta F$ over the quenched disorder should follow
\begin{equation}
\xi_{\Delta F}=\frac{2\sigma N^{1/2}}{\left| \epsilon+T\beta\log N \right|},
\label{eq:xitheor}
\end{equation}
and consequently display a divergence at $N_{div}=\exp[-\epsilon/T\beta]$ and a minimum at $N_{min}=\exp[2-\epsilon/T\beta]$, diverging at large $N$ (cf. Fig. \ref{fig:selfav}B). Thus, we can expect $\Delta F$ to be representative of a typical realization of the disordered interactions if $N>N_{div}$ and $N\sim N_{min}$.

In Fig. \ref{fig:selfav}C it is plotted the value of $\xi_{\Delta F}$ at different temperatures as a function of the length $N$ of the chain in semi--log scale, calculated over 500 realizations of the random interactions. For each temperature we show the points up to the largest value of $N$ for which we can guarantee the correct equilibration of the simulated--tempering algorithm. In the studied range of $N$, the calculated $\xi_{\Delta F}$ is decreasing,  thus suggesting that $N_{div}<N<N_{min}$ . Moreover, already for $N>10$ the $\xi_{\Delta F}$ assumes small values, indicating that the standard error on $\Delta F$ is of the order of a few percent of the mean.  That is, except for very short chains, the average of $\Delta F$  over the stochastic interactions are representative of their typical values. A similar behaviour is observed for the gyration radius $R_g$ of the polymer (see Fig. \ref{fig:selfav}D). 

%%%%%%%%%%%%%%%%%%%%%%%%%%
\section{Scaling of the free energy associated with the looping probability}
\label{sect:pd}
%%%%%%%%%%%%%%%%%%%%%%%%%%

From the same simulations used to estimate the degree of self--averageness, we calculated the values of $\overline{\Delta F}$ as a function of $N$, in order to estimate its scaling properties. 

\begin{figure}
\includegraphics[width=\linewidth]{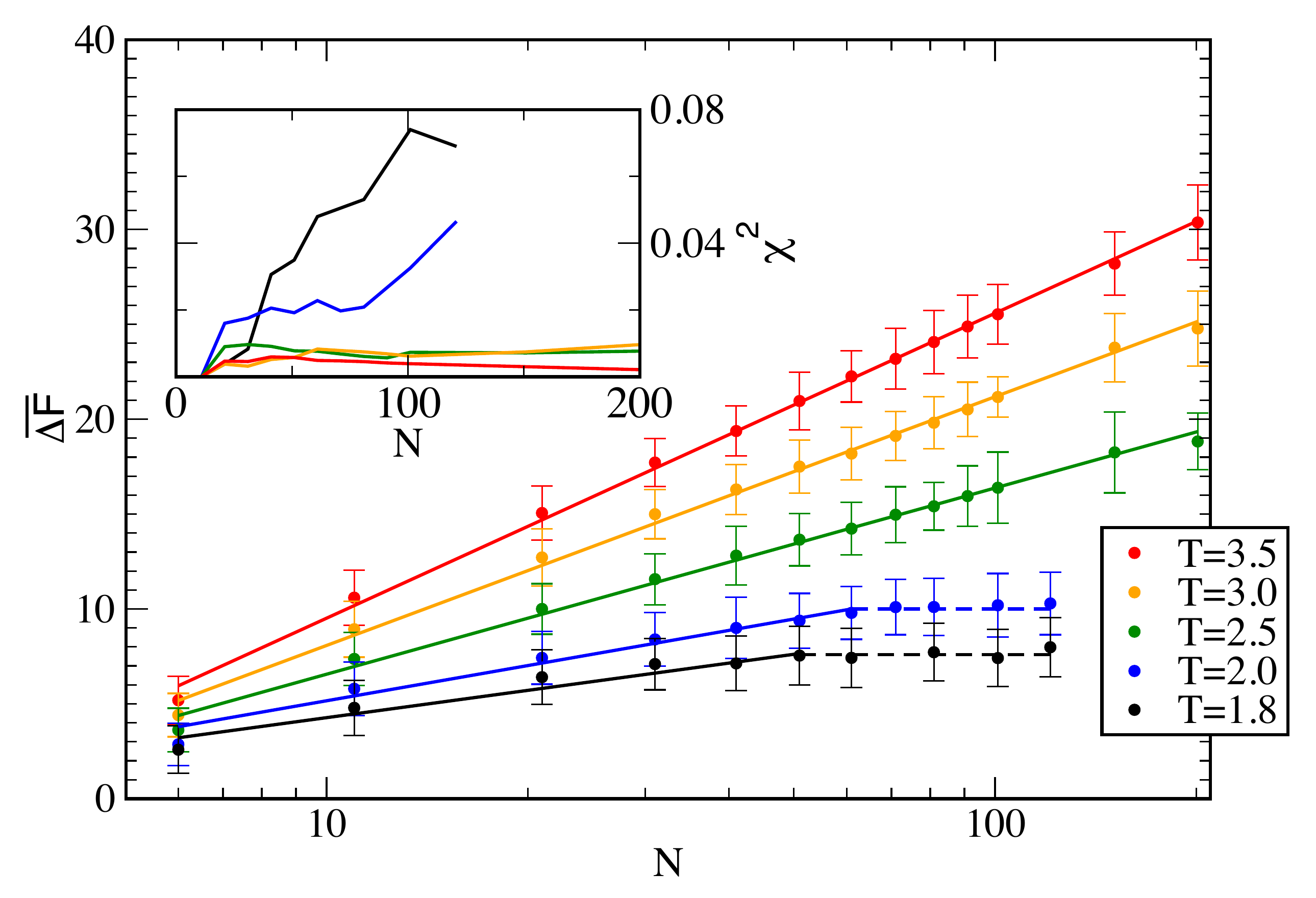}
\caption{(Color online) The average value of $\Delta F$  as a function of $N$, the latter displayed in a logarithmic scale. For each value of $N$, 500 realizations of the disordered interaction are simulated. The points are fitted according to Eq. \protect\ref{eq:beta}, and the corresponding line is drawn in the figure. In the inset, the $\chi^2$ associated with the fits calculated up to length $N$.}
\label{fig:fits}
\end{figure}

The linear fit of  $\Delta F$ as a function of $\log N$ is displayed in Fig. \ref{fig:fits} for simulations carried out at different temperatures. The linear fit appears good at $T>2.0$ and seem to worsen at lower temperatures. In particular, at $T\leq 2.0$ a power--law behaviour applies up to $N\approx 60$, while $\overline{\Delta F}$ appears weakly dependent on $N$ above $\approx 60$, similarly to the behaviour of a collapsed globule in a homopolymer.

Interpreting Eq. (\ref{eq:pf}) as the likelihood of observing a value of $\Delta F$ in a chain of specified length, the quality of the linear fit can be expressed in terms of the average log--likelihood, that is nothing else but
\begin{equation}
\chi^2=\frac{1}{Z_N} \sum_n^N \frac{ ( \overline{\Delta F}(n) - \epsilon - T\beta\log n )^2 }{ n\sigma^2 },
\label{eq:chi2}
\end{equation}
where $N$ is the length of the longest chain considered in the fit and $Z_N=\sum_n^N(n\sigma^2)^{-1}$. The values of $\chi^2$ as a function of $N$ are reported in the inset of Fig. \ref{fig:fits}. The fits of the points at $T>2.0$ display a constant or decreasing $\chi^2$ of the order of $10^{-2}$, while at lower temperatures it increases with $N$. However, even at low temperatures the value of $\chi^2$ remains lower than 1 for all the $N$ studied, indicating that the fitting line matches the points within their error bars. 

This is a result of the fact that both the estimation of $\beta$ and the quantification $\chi^2$ of the error of the fit  emphasize smaller polymers becuase for them the variability of $\Delta F$ due to the disordered interactions is smaller. In the case of longer polymers, $\overline{\Delta F}$ seems to become independent on $N$, but at the same time it becomes  less and less representative of a typical heteropolymer.  In fact, even if $\overline{\Delta F}$ were constant at large $N$, the leading term of Eq. (\ref{eq:chi2}) would be $\chi^2\sim N^{-1}\sum_n \log^2 n/n$; approximating the sum with an integral gives $\chi^2\sim \log^3{N}/N$ which vanishes at large $N$. In other words, it is the small-$N$ slope that determines $\beta$, because at large $N$ the free energy is dominated by the disorder. If the small--$N$ scaling properties are due to finite--size effects, these will thus dominate the results even when considering longer chains.

\begin{figure}
\includegraphics[width=\linewidth]{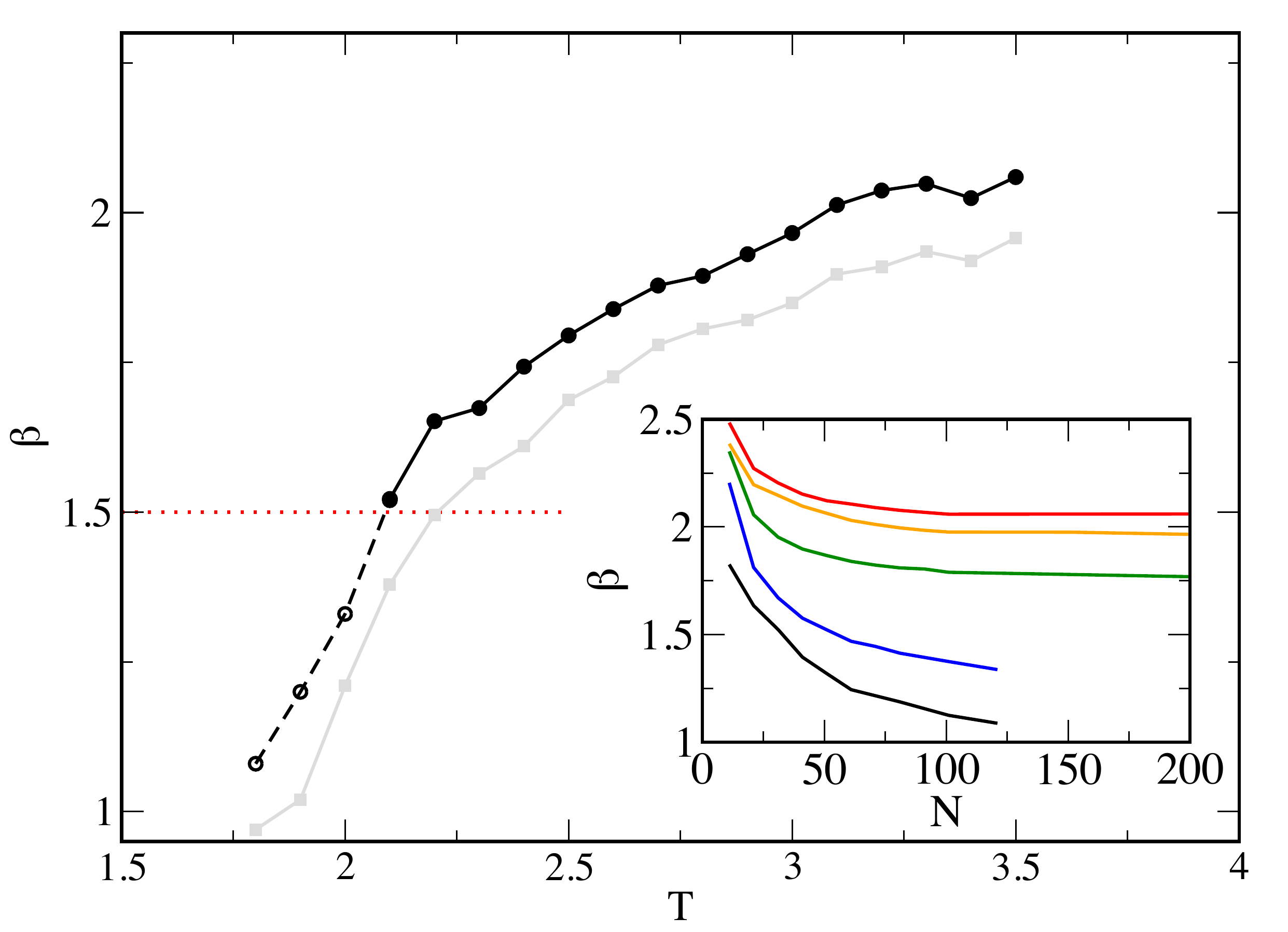}
\caption{(Color online) The exponents $\beta$ obtained using Eq.  \protect\ref{eq:beta}  at different temperatures from the fits of the simulated data up to the largest polymer we could equilibrate (circles). As a reference, the dotted curve indicates the exponent $3/2$ expected for an ideal chain. Empty circles indicate the exponents below the $\theta$--point, strongly affected by finite--size effects. The gray squares indicate the exponents found in a fit of $\log c$ versus $\log N$. In the inset, the exponent calculated from fits up to length $N$.}
\label{fig:exponents}
\end{figure}

The values of the parameter $\beta$ obtained from the fits at different temperatures are reported as solid circles in Fig. \ref{fig:exponents}. At high temperature ($T=3.5$) the scaling exponent $\beta$ converges to $2.06$, which is comparable with the value $2.10\pm 0.15$ obtained numerically for self--avoiding walks in three dimensions \cite{Guttmann:1973va}, and somewhat larger than the theoretical result $9/5$ obtained by de Gennes solving a zero--dimensional Ising model \cite{deGennes:book}.

\begin{figure}
\includegraphics[width=\linewidth]{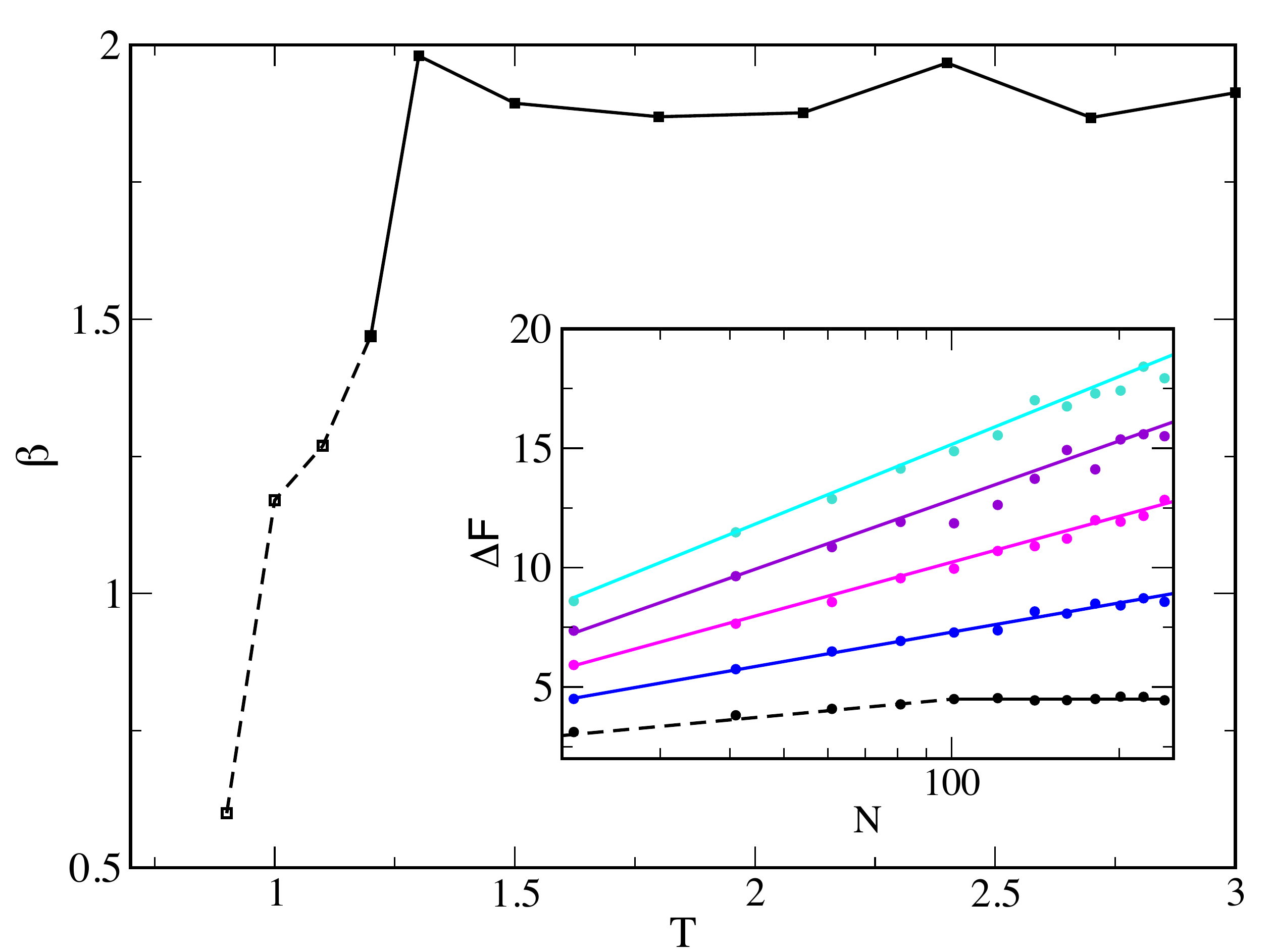}
\caption{(Color online) The scaling exponent $\beta$ calculated for a homopolymer (i.e., $\epsilon_0=-0.1$, $\sigma=0$) as a function of temperature $T$. Empty symbols indicate the exponents associated with finite--size behaviour (cf. dashed line in the inset). In the inset, the binding free energies whose fits were used to obtain the scaling exponents (the different sets correspond, starting from above, to $T=2.1$, $T=1.8$, $T=1.5$, $T=1.2$ and $T=0.9$).}
\label{fig:homo}
\end{figure}

As the temperature is decreased, $\beta$ decreases continuously to the value $\beta=3/2$ typical of the $\theta$--point at $T\approx 2.0$. This plot is markedly different from that of a homopolymer, in which case only two kinds of exponents are expected, associated with the coil state and the ideal behaviour at the $\theta$--point. In fact, the exponents found from numerical simulations of homopolymers of comparable size are displayed in Fig. \ref{fig:homo}.  Moreover, even a random heteropolymer in the coil-- or $\theta$--state in the limit of short interaction range is expected to display the same exponents of the homopolymer, superposed to an exponential cutoff \cite{Tiana:2015ir}.

Below the $\theta$--point the fit gives exponents $1\lesssim \beta\lesssim 1.5$ (cf. empty circles in Fig. \ref{fig:exponents}). Since the small--$N$ contribution dominates due to the dependence on $N$ of the denominator at the exponent of Eq. (\ref{eq:pf}),  the exponents $\beta$ seem to converge to a $N$--independent value, different from zero, even below the $\theta$--point (cf. inset of Fig. \ref{fig:exponents}). 

The scaling of $\overline{\Delta F}$  below the $\theta$--point with exponents lower than $3/2$ is a finite--size effect, also present in homopolymers (cf. Fig. \ref{fig:homo}). This is a consequence of the fact that if the polymer is too short, it is not able to define a bulk volume, necessary for the looping entropy to lose its dependence on $N$,  but its volume essentially coincides with its surface. The order of magnitude of $N$ below which this effect takes place is found by $4\pi R^2\cdot 2r_H=4/3\pi R^3$, with $R=r_HN^{1/3}$ in a globule, that is $N= 6^3\approx 10^2$, in agreement with what shown in Fig.  \ref{fig:homo}.

Often a simple regression of $\log c$ versus $\log N$ was applied to the analysis of the scaling properties of the contact probability \cite{Buscaglia:2006} of biopolymers. This is more difficult to justify theoretically than the fit described in Sect. \ref{sect:theo}. Anyway, the result of such a fit are displayed with gray squares in Fig. \ref{fig:exponents}. The resulting exponents are slightly smaller than those obtained with the two--state model described above, but in this case the (unweighted) $\chi^2$ of the fit ranges from 0.2 at high temperature to $\approx 1.8$ at low temperature. At variance with the the weighted fit described above, in this case the $\chi^2$ of the fit, as well as the value of the exponents, depend on the specific range of $N$ employed in the simulations.

%%%%%%%%%%%%%%%%%%%%%%%%%%%
\section{Compactness of the polymer}
%%%%%%%%%%%%%%%%%%%%%%%%%%%

In order to compare the exponents $\beta$ found for the random heteropolymer with those known from the theory of homopolymers, it is interesting to understand whether the polymer is, at the different temperatures studied above, in a globular or in a coil state. This problem is well--defined because the thermal average $R_g$ of the gyration radius results to be self--averaging (see Sect. \ref{sect:selfav}), and consequently we can study its average $\overline{R_g}$ over the realizations of the disordered interaction. On the other hand, it is complicated by the small size of the system, while a globule--coil phase transition is defined, strictly speaking, only for an infinitely--long polymer.

\begin{figure}
\includegraphics[width=\linewidth]{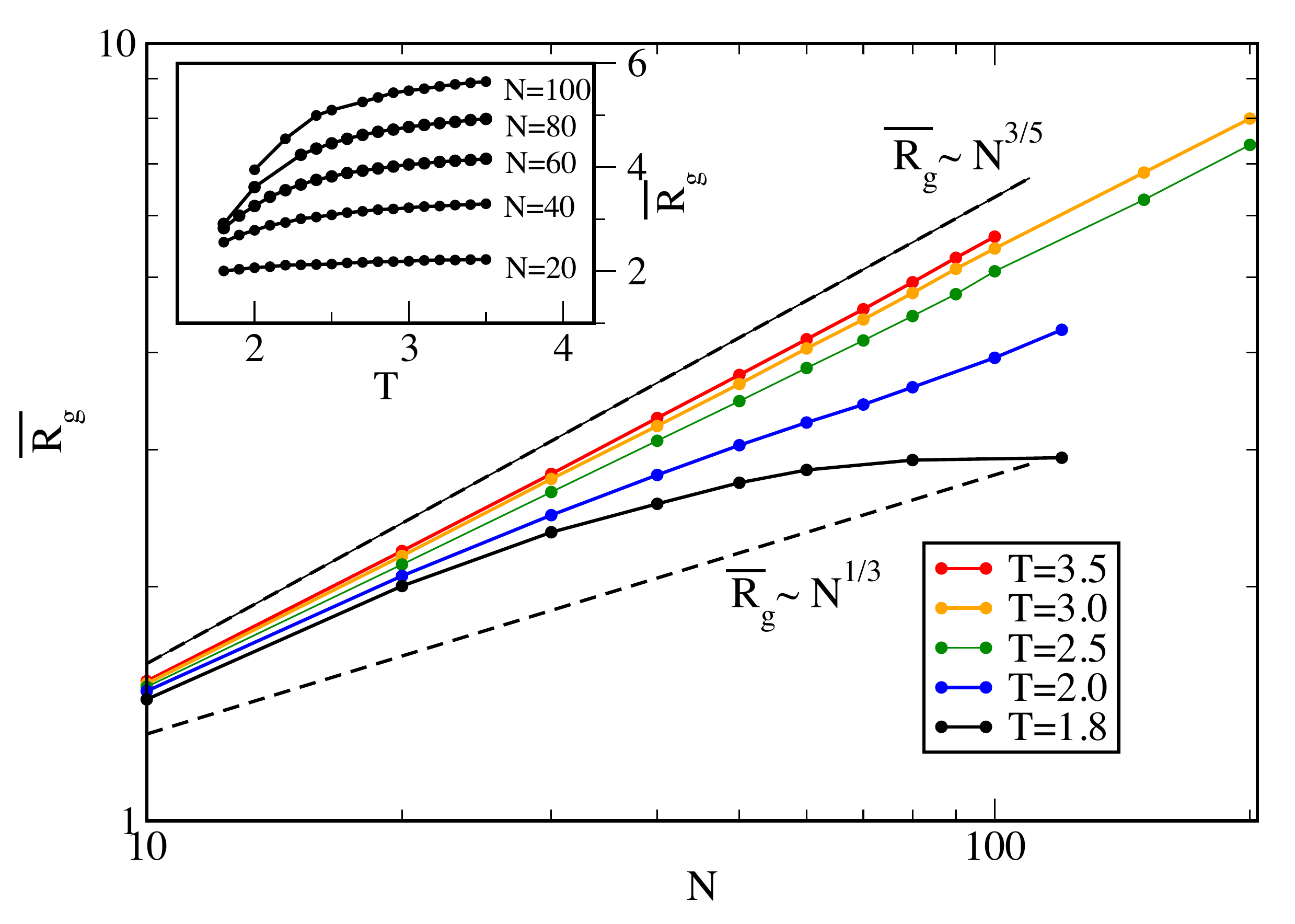}
\caption{(Color online) The average gyration radius $\overline{R_g}$ at different temperatures as a function of the length of the chain plotted in log--log scale. As a reference, we indicate with dashed lines the $N^{3/5}$ curve expected for a random coil and the $N^{1/3}$ curve expected for a globule. In the inset, the value of $R_g$ as a function of temperature for different lengths $N$. }
\label{fig:gyration}
\end{figure}

The average value of $\overline{R_g}$ as a function of $N$ is displayed in log--log scale in Fig. \ref{fig:gyration} at different temperatures. For $T\geq 3.0$ the curves overlap almost perfectly to each other, with a slope of $\approx 3/5$,  that of a random coil in the case of a homopolymer. This is not unexpected, since at high temperature the heterogeneity in the interactions within the chain becomes negligible with respect to $T$, and the heteropolymer behaves effectively as a homopolymer.

For temperatures $T<3.0$ the slope of $\log \overline{R_g}$ versus $\log N$ decreases and reaches $1/2$, the value that homopolymers display at the $\theta$--point, at $T\approx 2.1$. If one decreases the temperature further, the curve is no longer linear in the range of $N$ under consideration. This is likely to be a finite--size effect, since the gyration radius has to grow at least as $N^{1/3}$, corresponding to a fully compact structure.

The decrease of $\overline{R_g}$ as a function of $T$ can also be visualized directly in the inset of Fig. \ref{fig:gyration} for each value of $N$. A clear transition in $\overline{R_g}$ cannot be seen at any value of $N$. At large values of $N$, where transitions are expected to be sharper, we are not able to equilibrate the lowest temperatures, corresponding to the compact phase. Consequently, we are not able to highlight clearly a globule--coil transition, similar to that of homopolymers.

The clearest set of data is that calculated for $N=60$. At $T=1.8$ the mean gyration radius is $2.7$, not far from that of a maximally--compact globule, that is $N^{1/3}\cdot r_H=2.4$. At $T=2.0$ the value of $\overline{R_g}$ is $3.2$, close to that associated with that of an ideal chain, that is $0.41\cdot N^{1/2}=3.18$. Anyway, the curve increases smoothly from the more compact to the more elongated conformations.

Summing up, the random heteropolymer displays at high temperature  properties of the radius of gyration similar to those of homopolymers, including a $\theta$--point at which the size of the heteropolymer scales as that of an ideal chain. A lower temperatures, in the range of lengths we could equilibrate, the size is dominated by finite--size effects.

%%%%%%%%%%%%%%%%%%%%%%%%%%%
\section{Scaling properties within a fixed--length chain}
%%%%%%%%%%%%%%%%%%%%%%%%%%%

Sometimes the experimental data to analyze is not the looping probability of polymers of different lengths, but the looping probabilities of the various segments, of different lengths, within a given polymer. This is, for example, the case of chromosome conformation capture experiments on the chromatin fibre \cite{Lieberman:2009}. The standard way of extracting the scaling exponent is a linear regression of $\log c(i,j)$ versus $\log |i-j|$ of the whole set of data, where $|i-j|\leq N$ is the length of the segment starting at monomer $i$ and ending at monomer $j$ of the $N$--bead polymer. It was also suggested that fitting $c$ versus $n$ is a better strategy \cite{Clauset:2009}; this is however unwise in the case of heteropolymers, because of the lack of self--averaging of $c$ (cf. Sect. \ref{sect:selfav}).

Anyway, if the heterogeneity in the looping probability at fixed inter--monomer linear distance is due to the variability of the interactions, the correct way of extracting the scaling behaviour is similar to that described in Sect. \ref{sect:theo}.
As in the case of heteropolymers of different lengths, one can define a looping free energy $\Delta F$ (cf. Eq. (\ref{eq:df})) and develop calculations similar to those which lead to Eq. (\ref{eq:beta}). However, now Eq. (\ref{eq:cN}) depends on $|i-j|$ instead of $N$, that is
\begin{equation}
\Delta F(i,j)=\Delta E+\epsilon+T\beta'\log |i-j|,
\label{eq:cN2}
\end{equation}
where we define the scaling exponent as $\beta'$ to distinguish it from that of varying--size polymers. Now  Eq. (\ref{eq:gauss}) is still valid, but $N$ is fixed. 
The result is that, according to this model, $\beta'$ should be obtained by an unweighted linear regression of $\Delta F(i,j)$ versus $\log |j-i|$. Here, the main difference with Eq. (\ref{eq:beta}) is the lack of weights in the sum. 

As one is usually interested in the scaling properties of any two monomers as a function of their distance $n$ along the chain, and not of two specific monomers $i$ and $j$ (which is, anyway, hardly self--averaging), a more convenient quantity to study is  $\Delta F(n)=(N-n+1)^{-1}\sum_j \Delta F(j,j+n)$. From the properties of convolutions of Gaussian distributions, from Eq. (\ref{eq:pf}) one obtains
\begin{align}
p(\Delta F(n))&=\sqrt{   \frac{(N-n+1)}{4\pi N\sigma^2}   } \nonumber\\   
&\exp\left(   -\frac{\left(\Delta F - T\beta\log n +\epsilon \right)^2}   {4N(N-n+1)^{-1}\sigma^2}   \right).
\label{eq:pf2}
\end{align}
Consequently, $\beta'$ can be found, in analogy with Eq. (\ref{eq:pf}), from a linear fit of $\Delta F(n)$ versus $\log n$, weighted by $(N-n+1)/N$. Operatively, this is not different from a linear regression of $\Delta F(i,j)$ versus  $\log |i-j|$, since $(N-n+1)$ is just the multiplicity of pairs of monomers at linear distance $n$.

\begin{figure}
\includegraphics[width=\linewidth]{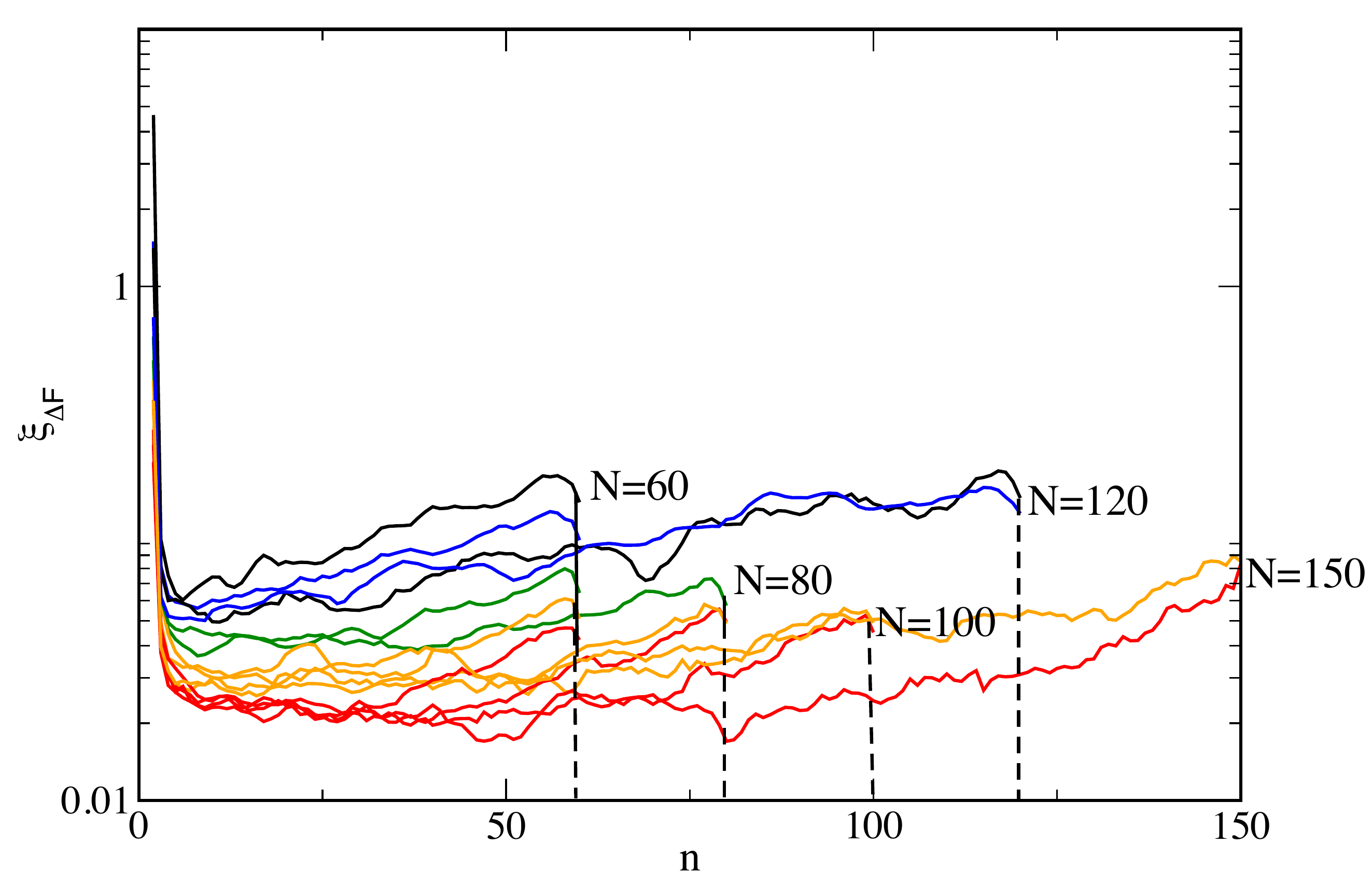}
\caption{(Color online) The degree of self--averaging of $\Delta F(n)$ calculated at different values of $N$ and of the temperature. The color code indicates the temperature and is the same as in Fig, \protect\ref{fig:selfav}}
\label{fig:selfav_fixedN}
\end{figure}

The parameter $\xi_{\Delta F(n)}^2$ which describe the degree of self--averaging of $F(n)$ is displayed in Fig. \ref{fig:selfav_fixedN}. For each $T$ and $N$ it displays a non--monotonic behaviour as a function of $n$. At low $n$, $\xi_{\Delta F(n)}^2$ is large as in the case of fixed--length heteropolymer (cf. Fig. \ref{fig:selfav}); then it drops because each value of $\Delta F(n)$ is the average not only on the realizations of the disorder, but also on the $N-n+1$ segments of length $n$, and each of them can be regarded as a realization of the disorder as well (see the discussion in ref. \cite{Tiana:2015ir}). As $n$ increases, this effect diminishes, and $\xi_{\Delta F(n)}^2$ increases. For fixed $n$, $\xi_{\Delta F(n)}^2$ displays at each temperature in the region $n\sim N$ a decreasing behaviour, which suggests the self--averaging character of this quantity.

\begin{figure}
\includegraphics[width=\linewidth]{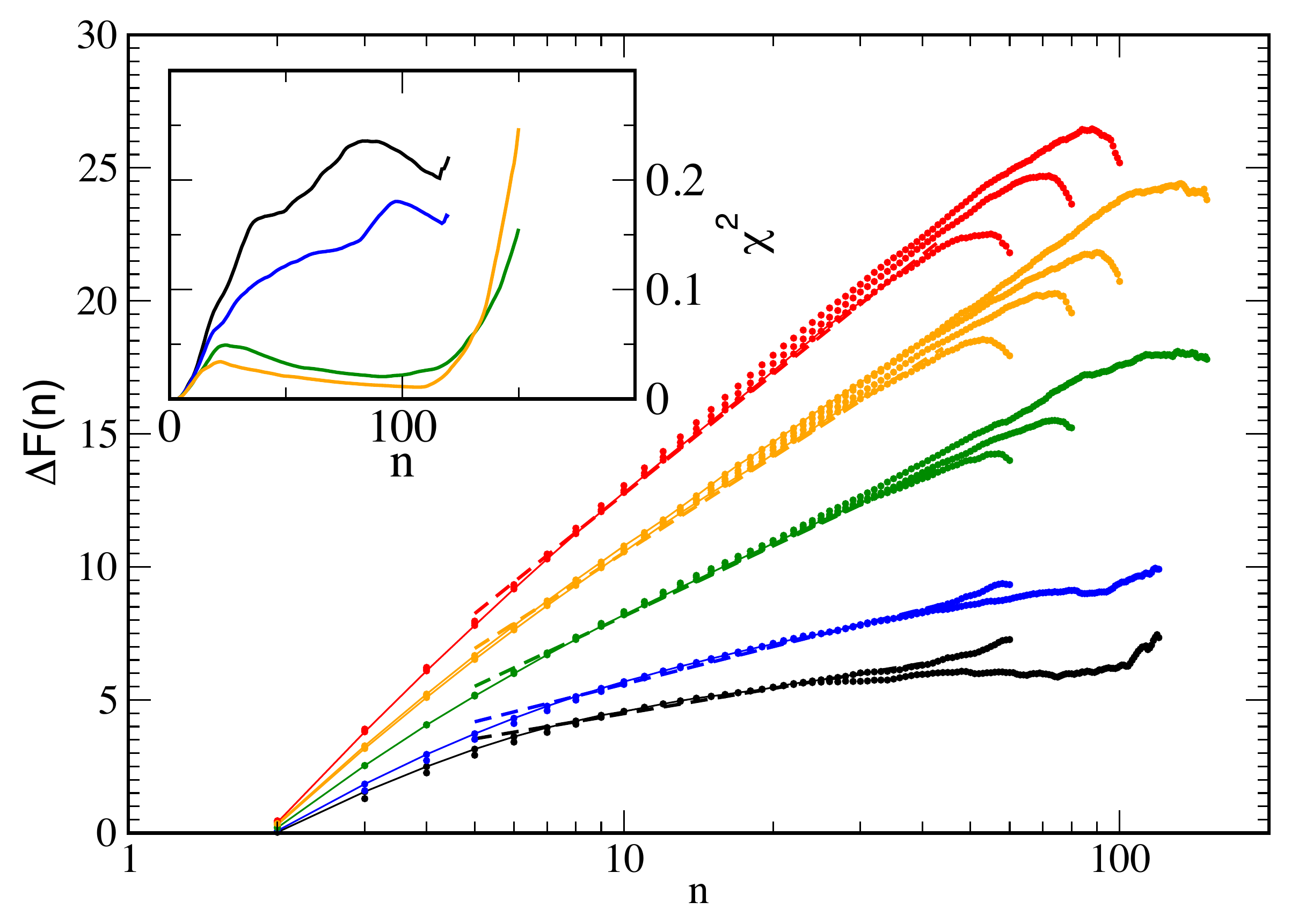}
\caption{(Color online) The scaling of $\overline{\Delta F(n)}$ as a function of $\log n$ at different temperatures (color code of Fig. \protect\ref{fig:selfav}) for different values of $N$. The fit, done between $N=6$ and $n=60$, is displayed with a dashed line. In the inset, the $\chi^2$ associated with the fit up to length $n$.}
\label{fig:fit_fixedN}
\end{figure}

The behavior of $\overline{\Delta F(n)}$ as a function of $\log n$ is displayed in Fig. \ref{fig:fit_fixedN}, obtained from polymers with $N=60,\,80,\,100,\,120$ at different temperatures. The $\chi^2$, weighted according to Eq. (\ref{eq:pf2}), associated with the fit from $n=6$ (below which self--averaging is absent, cf. Fig. \ref{fig:selfav_fixedN}) to varying $n$ is displayed in the inset of Fig. \ref{fig:fit_fixedN}. At $T>2.0$, corresponding to the elongated phase of the polymer (cf. previous section), the linear fit is very good  except when $n\approx N$. At lower temperatures, only the central region is linear ($6\lesssim n\lesssim60$), while for $n\sim N$ the curve bends down similarly to what expected for a homopolymeric globule. However, in all cases the associated  $\chi^2$ remains lower than 1, due to the larger weight of small $n$ to the fit. 

\begin{figure}
\includegraphics[width=\linewidth]{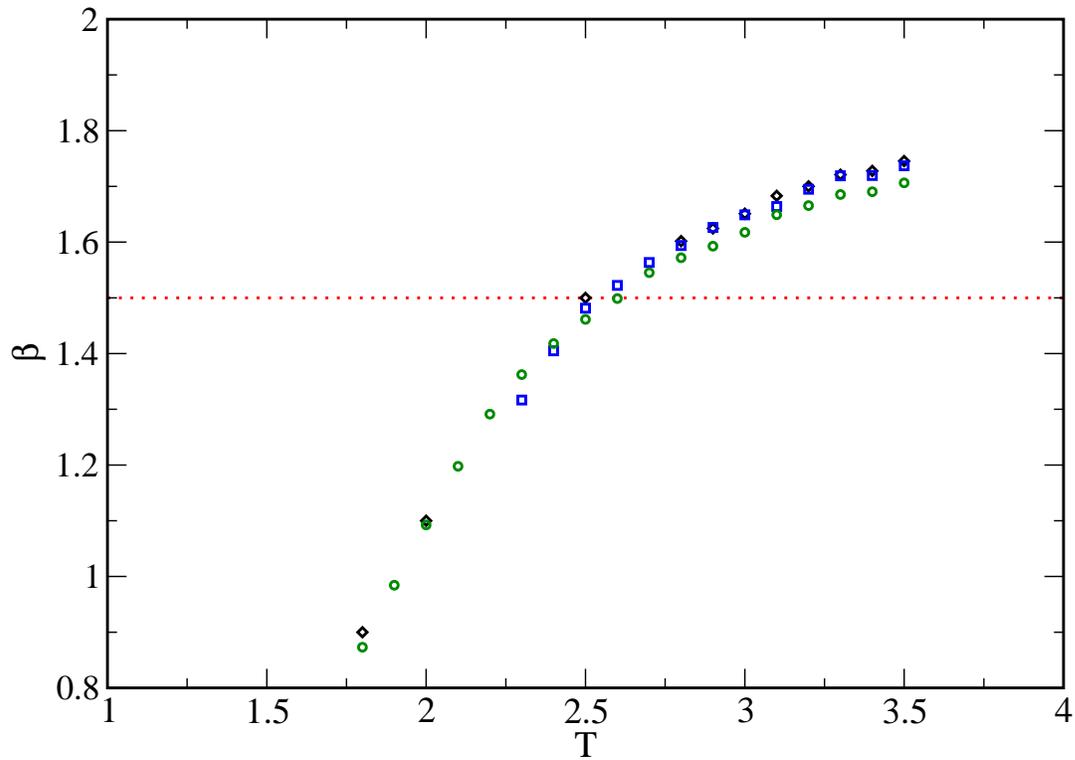}
\caption{(Color online) The exponents $\beta'$ associated with the fits of $\Delta F(n)$ versus $n$ (solid black symbols), for the case $N=60$ (green circles), $N=80$ (blue squares) and $N=120$ (black diamonds). }
\label{fig:exponent_fixedN}
\end{figure}

The values of $\beta'$ obtained from the fits is displayed in Fig. \ref{fig:exponent_fixedN}.  Overall, the values of $\beta'$ are smaller than those of $\beta$ corresponding to the same temperature. At the highest temperature it displays the value $\approx 9/5$ predicted for self--avoiding walks. At low temperatures, $\beta'$ can reach values as low as $0.92$. The reason is again that finite--size effects are amplified by the larger weight of small fragments of the chain, which is anyway unavoidable because fragments with $n\sim N$ are dominated by disorder. 

%%%%%%%%%%%
\section{Discussion and Conclusions}
%%%%%%%%%%%

The free energy difference between looped and unlooped states within a two--state model provides a consistent way of studying the scaling properties associated with the  looping mechanism with respect to the length of the random heteropolymer. From a theoretical argument and from numerical simulations, based on a self--adjusting simulated tempering technique, the fluctuations about the average over the realizations of the random interaction within the heteropolymer are small, in the range of length of the order of $10^2$ monomers but not in the thermodynamic limit.

Polymers of $\sim 10^2$ monomers are the longest systems we could guarantee equilibration, although with a consistent computational effort. Fortunately, this is the typical size of biological polymers. In fact, protein domains have an average length of 150 residues \cite{Nussinov:1998}. Topological associating domains in mammalian chromatin display a typical length of $10^6$ bases, corresponding to $10^2$ Kuhn lengths \cite{Dekker:2008}.

At high temperature, where the polymer is elongated, the looping probability of random heteropolymers displays a scaling exponent which varies continuously with respect to the temperature from $\approx 2.05$ to $1.5$. This is different from the behaviour of homopolymers, for which only two possible exponents are expected. 

At lower temperatures, corresponding to a compact phase of the heteropolymer, the determination of the scaling exponent is more cumbersome. Short chains display significant finite--size effects, resulting in a scaling of the looping probability with exponents smaller than 1.5. Longer chains display large disorder--dependent variability, which down--weights the determination of the exponent and the evaluation of the associated error. This amplifies the role of finite--size effects in the determination of the exponents even of large chains.

This phenomenon operates, for different reasons, both when considering chains of different lengths and segments of different lengths in a fixed--length heteropolymer. In the former case, the looping free energy is affected by the disorder provided by the internal energy of the chain, which is an extensive quantity. In the latter case, the free energy must be averaged over all the segments of the same length to be self--averaging, and the number of such segments decreases with the overall length of the chain. Anyway, fits of self--averaging free energies at low temperatures emphasize finite--size effects, resulting in exponents smaller than $3/2$.  

In the study of looping probability of the chromatin fibre, it is quite common to obtain scaling exponents lower than those which are typical of homopolymers. While out--of--equilibrium effects \cite{Mirny:2011}, particle--mediated interactions \cite{Barbieri:2012} or energy--driven mechanisms \cite{Goloborodko:2015,Sanborn:2015} have been advocated so far to explain such small exponents, the calculations described above suggest that finite--size effects, combined with the heterogeneity of the interactions in the chain, are sufficient to justify the experimental data. Of course the present model does not provide a mechanistic interpretation of the observed exponents, but suggests that scaling exponents cannot be the only quantitative observable used to build and validate a model.  

The values of $\beta$ found in the variable--length segments of a fixed--length chain result smaller than those of a set of chains of different lengths. There are two differences between the two cases. The former is that considering the variable--lengths segments of the same chain leaves correlations in the contact energies, which are absent when considering different realizations of varying-length chains. Moreover, when studying the variable--lengths segments of the same chain, the "tails" of the chain (i.e., the segments $1$ to $i-1$ and $j+1$ to $N$, when studying the looping of $i$ with $j$) may play a role. As a matter of fact, also for homopolymers it was shown\cite{Chan:1989un} that the length of the tail can affect considerably the looping mechanism. The reason is that the excluded volume of the tail can shield the two monomers defining the loop, decreasing their binding probability.

\begin{figure}
\includegraphics[width=\linewidth]{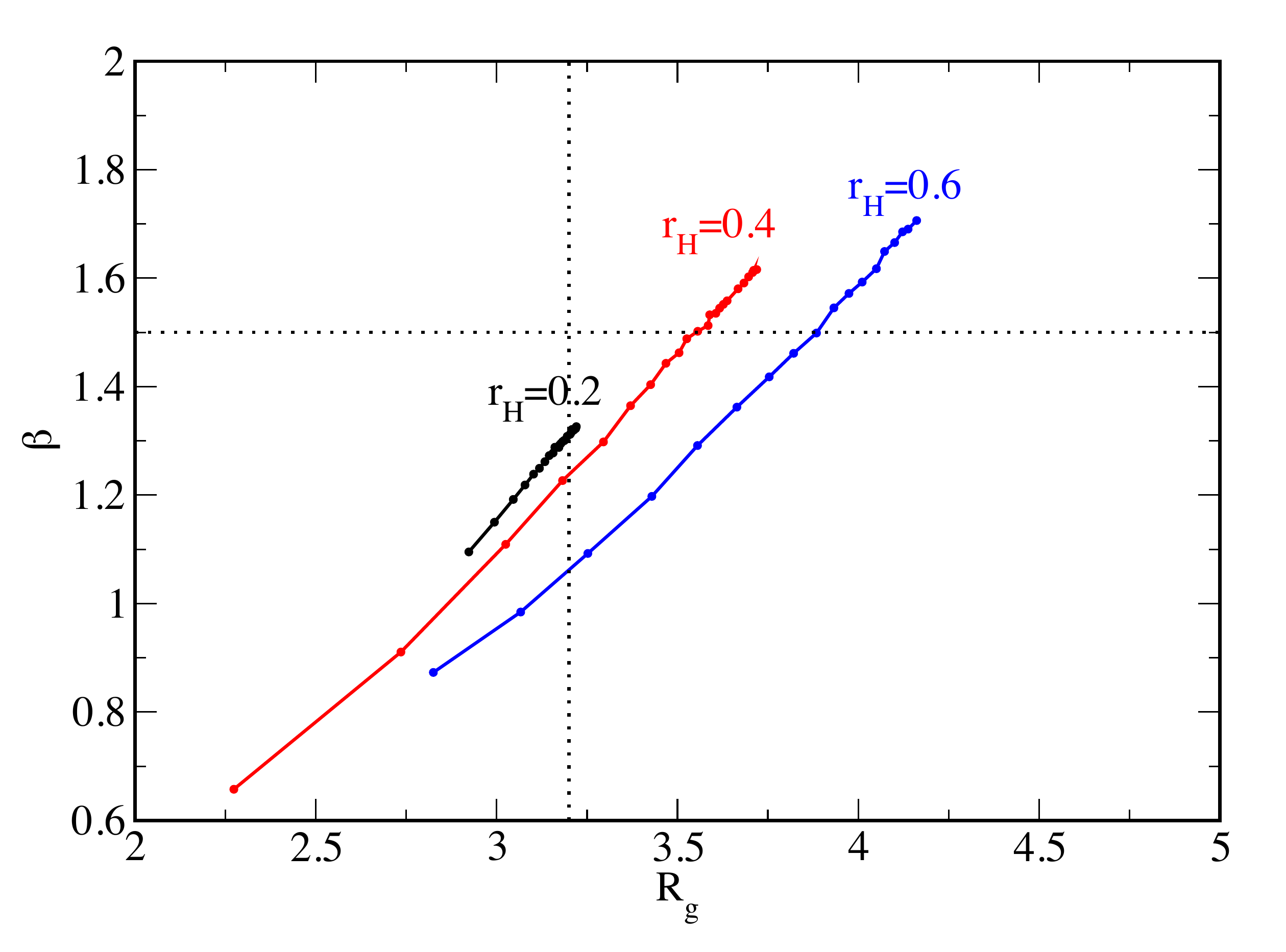}
\caption{(Color online) The exponents $\beta$ found at different temperatures, corresponding to different gyration radii $R_g$, using models with different length scale of the interaction potentials. The segments of a chain with $N=60$ are used to calculate the values of $\beta$. The dotted lines indicates the expected value of $R_g$ and of $\beta$ at the $\theta$--point.}
\label{fig:rhc}
\end{figure}

To investigate this point, we have repeated the simulations with different potentials, defined by different choices of the hardcore radius $r_{HC}$ (and interaction radius proportional to $r_{HC}$), calculating the value of the exponent $\beta$ for each of them. In Fig. \ref{fig:rhc} we show the result of these calculations. Since  models with different $r_{HC}$ display different temperature scales for the coil--globule transition, we use as independent variable the gyration radius $R_g$. For each value of $R_g$, decreasing $r_{HC}$ the resulting $\beta$ increases towards the values found with chains of different lengths, suggesting that the shielding effect plays a role in determining the difference between the two cases.

These results also suggests that the difference between the present numerical calculations and the analytical results found in ref. \cite{Tiana:2015ir}, namely that for $T\geq\theta$ the exponent of a heteropolymer should not change with respect to the homopolymeric case, while only an exponential cutoff appears in the looping probability, can be associated with the hypothesis $r_{HC}\to 0$ used in the analytical calculations.

%%%%%%%%%%%%%%

\newpage

\begin{thebibliography}{99}
%%%%%%%%%%%%%%
\bibitem{Blackwood:1998} E. M. Blackwood and J. T. Kadonaga, Science {\bf 281}, 60 (1998)
\bibitem{Bruun:2010} S. W. Bruun, V.  Iesmantavicius, J. Danielsson, and F. M. Poulsen, Proc. Natl. Acad. Sci. USA {\bf 107}, 13306
\bibitem{Lieberman:2009} E. Lieberman-Aiden, N. L. van Berkum, L. Williams, M. Imakaev, T. Ragoczy, A. Telling, I. Amit, B. R. Lajoie, P. J. Sabo, M. O. Dorschner, R. Sandstrom, B. Bernstein, M. A. Bender, M. Groudine, A. Gnirke, J. Stamatoyannopoulos, L. A. Mirny, E. S. Lander, and J. Dekker, Science {\bf 326}, 289 (2009)
\bibitem{Sanborn:2015} A. L. Sanborn, S. S. P. Rao, S.-C. Huanga, N. C. Duranda, M. H. Huntley, A. I. Jewett, I. D. Bochkova, D. Chinnappan, A. Cutkosky, J. Li, Kristopher P. Geeting, A. Gnirkee, A. Melnikove, D. McKenna, E. K. Stamenova, E. S. Lander, and Erez Lieberman Aiden, Proc. Natl. Acad. Sci USA {\bf 112}, E6456 (2015)
\bibitem{Buscaglia:2006} M. Buscaglia, L. J. Lapidus, W. A. Eaton and J. Hofrichter, Biophys. J. {\bf 91}, 276 (2006)
\bibitem{Plaxco:1998} K. W. Plaxco, K. T. Simons, and D. Baker, J. Mol. Biol. {\bf 277}, 985 (1998)
\bibitem{Liu:2006} L. Liu and C. Hyeon, arXiv:1604.00472
\bibitem{deGennes:book} P.--G. de Gennes, {\it Scaling Concepts in Polymer Physics}, Cornell University Press, 1979
\bibitem{Mirny:2011} L. Mirny, Chromosome Res. {\bf 19}, 37 (2011)
\bibitem{Grosberg:1988} A. Yu. Grosberg, S. K. Nechaev and E. I. Shakhnovich, J. Phys. France {\bf 49}, 2095 (1988)
\bibitem{Goloborodko:2015} Goloborodko, J. F. Marko and L. A. Mirny, bioRxiv:10.1101/021642
\bibitem{Barbieri:2012} M. Barbieri, M. Chotalia, J. Fraser, L.--M. Lavitas, J. Dostie, A. Pombo, and M. Nicodemi, Proc. Natl. Acad. Sci. USA {\bf 109}, 16173 (2011)
\bibitem{Lifshits:1942} I. M. Lifshits, Zh. Eksp. Teor. Fiz. {\bf 12}, 117 (1942)
\bibitem{Swendsen:1986} R. H. Swendsen and J.--S. Wang, Phys. Rev. Lett. {\bf 57}, 2607 (1986)
\bibitem{Marinari:1992} E. Marinari and G. Parisi, Europhys. Lett. {\bf 19}, 451 (1992)
\bibitem{Tiana:2011} G. Tiana and L. Sutto, Phys. Rev. E {\bf 84}, 061910 (2011)
\bibitem{Giorgetti:2014} L. Giorgetti, R. Galupa, E. P. Nora, T. Piolot, F. Lam, J. Dekker, G. Tiana and E. Heard,  Cell, {\bf 157}, 950 (2014)
\bibitem{Brout:1959} R. Brout, Phys. Rev. {\bf 115}, 824 (1959)
\bibitem{Tiana:2015ir} G. Tiana, Phys. Rev. E. {\bf 92}, 010702R (2015) 
\bibitem{Shakhnovich:1989} E. I. Shakhnovich and A. M. Gutin, J. Phys. France {\bf 50}, 1843 (1989)
\bibitem{Derrida:1981uc} B. Derrida, Phys. Rev. B. {\bf 24}, 2613 (1981)
\bibitem{Guttmann:1973va} A. J. Guttman and M. F. Sykes, J. Phys. C {\bf 6}, 945 (1973)
\bibitem{Clauset:2009} A. Clauset, C S. Shalizi and M. E. J. Newman, SIAM Rev. {\bf 51}, 661 (2009)
\bibitem{Chan:1989un} H. S. Chan and K. A. Dill, J. Chem. Phys. {\bf 90}, 492 (1989)
\bibitem{Nussinov:1998} D. Xu and R. Nussinov, Folding \& Design {\bf 3}, 11 (1998)
\bibitem{Dekker:2008} J. Dekker, J. biol. Chem. {\bf 283}, 34532 (2008)
\end{thebibliography}
\end{document}